\documentclass[twocolumn]{aastex631}
\usepackage{CJK}
\usepackage{multirow}
\usepackage{makecell}
\usepackage{graphicx}
\newcommand{\hii}{\mbox{H\,\textsc{ii}}}
\newcommand{\hi}{\mbox{H\,\textsc{i}}}
\begin{document}
\begin{CJK*}{UTF8}{gbsn}
\title{East Asian VLBI Network astrometry toward the star-forming region G040.96+02.48 in the Extreme Outer Galaxy}

\shorttitle{EAVN observations of G040.96+02.48}
\shortauthors{Shen et al.}

\author{Xianjin Shen}
\affiliation{Purple Mountain Observatory, Chinese Academy of Sciences, Nanjing 210008, People's Republic of China}
\affiliation{School of Astronomy and Space Science, University of Science and Technology of China, Hefei 230026, People's Republic of China}

\author{Zehao Lin}
\affiliation{Purple Mountain Observatory, Chinese Academy of Sciences, Nanjing 210008, People's Republic of China}
\author{Nobuyuki Sakai}
\affiliation{National Astronomical Research Institute of Thailand (Public Organization), 260 Moo 4, T. Donkaew,
A. Maerim, Chiang Mai, 50180, Thailand}
\affiliation{Mizusawa VLBI Observatory, National Astronomical Observatory of Japan, 2-12 Hoshigaoka, Mizusawa, Oshu, Iwate 023-0861, Japan}
\author{Ye Xu}
\affiliation{Purple Mountain Observatory, Chinese Academy of Sciences, Nanjing 210008, People's Republic of China}
\affiliation{School of Astronomy and Space Science, University of Science and Technology of China, Hefei 230026, People's Republic of China}
\author{Shuaibo Bian}
\affiliation{Purple Mountain Observatory, Chinese Academy of Sciences, Nanjing 210008, People's Republic of China}
\author{Yuanwei Wu}
\affiliation{ National Time Service Center, Key Laboratory of Precise Positioning and Timing Technology, Chinese Academy of Sciences, Xi'an 710600, Peopleʼs Republic of
China}
\author{Yan Sun}
\affiliation{Purple Mountain Observatory, Chinese Academy of Sciences, Nanjing 210008, People's Republic of China}
\affiliation{School of Astronomy and Space Science, University of Science and Technology of China, Hefei 230026, People's Republic of China}
\author{Dejian Liu}
\affiliation{College of Mathematical and Physics, China Three Gorges University, Yichang 443002, China} 
\affiliation{Center for Astronomy and Space Sciences, China Three Gorges University, 443002, China}
\author{Jingjing Li}
\affiliation{Purple Mountain Observatory, Chinese Academy of Sciences, Nanjing 210008, People's Republic of China}
\affiliation{School of Astronomy and Space Science, University of Science and Technology of China, Hefei 230026, People's Republic of China}
\author{Bo Zhang}
\affiliation{Shanghai Astronomical Observatory, Chinese Academy of Sciences, Shanghai 200030, China}
\author{Shuangjing Xu}
\affiliation{Korea Astronomy $\&$ Space Science Institute, 776, Daedeokdae-ro, Yuseong-gu, Daejeon 34055, Republic of Korea}
\affiliation{Shanghai Astronomical Observatory, Chinese Academy of Sciences, Shanghai 200030, China}
\author{Tomoaki Oyama}
\affiliation{Mizusawa VLBI Observatory, National Astronomical Observatory of Japan, 2-12 Hoshigaoka, Mizusawa, Oshu, Iwate 023-0861, Japan}
\author{Chungsik Oh }
\affiliation{Korea Astronomy $\&$ Space Science Institute, 776, Daedeokdae-ro, Yuseong-gu, Daejeon 34055, Republic of Korea}
\author{Wu Jiang}
\affiliation{Shanghai Astronomical Observatory, Chinese Academy of Sciences, Shanghai 200030, China}
\author{Lang Cui}
\affiliation{Xinjiang Astronomical Observatory, Chinese Academy of Sciences, Urumqi 830011, China}
\author{Pengfei Jiang}
\affiliation{Xinjiang Astronomical Observatory, Chinese Academy of Sciences, Urumqi 830011, China}
\author{Guanghui Li}
\affiliation{Xinjiang Astronomical Observatory, Chinese Academy of Sciences, Urumqi 830011, China}
\author{Mareki Honma}
\affiliation{Mizusawa VLBI Observatory, National Astronomical Observatory of Japan, 2-12 Hoshigaoka, Mizusawa, Oshu, Iwate 023-0861, Japan}
\author{Se-Jin Oh}
\affiliation{Korea Astronomy $\&$ Space Science Institute, 776, Daedeokdae-ro, Yuseong-gu, Daejeon 34055, Republic of Korea}
\author{Zhi-Qiang Shen}
\affiliation{Shanghai Astronomical Observatory, Chinese Academy of Sciences, Shanghai 200030, China}
\author{Na Wang}
\affiliation{Xinjiang Astronomical Observatory, Chinese Academy of Sciences, Urumqi 830011, China}

\correspondingauthor{Zehao Lin}
\email{linzh@pmo.ac.cn}

\begin{abstract}
Accurate astrometric measurements for star-forming regions located on the far side of the Milky Way remain scarce.
In this work, we present the astrometric results for a 22\,GHz water maser associated with star-forming region G040.96+02.48 located on the far side of the Milky Way, using the East Asian VLBI Network. The target water maser's proper motion was determined to be ($\mu_{\alpha}\cos\delta, \mu_{\delta}$) = ($-2.06_{-0.51}^{+0.53}$,  $-2.95_{-0.44}^{+0.45}$)~mas~yr$^{-1}$. The derived three-dimensional kinematic distance to the star-forming region is 20.2$\pm$3.2\,kpc, placing it slightly outside the Outer Scutum$-$Centaurus Arm. The corresponding vertical height of 872$\pm$139\,pc indicates a significant warp of the outer Galactic disk, which is in good agreement with the latest precessing warp model. Moreover, the resulting peculiar motions reveal a complex kinematic pattern, characterized by a large outward radial velocity of $-32\pm$18\,km~s$^{-1}$.
Our observations substantially expand the valuable sample of star-forming regions with accurate astrometric measurements in the Extreme Outer Galaxy.
\end{abstract}


\section{Introduction} 
\label{sec:introduction}
It has long been established that the Milky Way is one of a large class of disk galaxies possessing large-scale structures, including a central bar \citep{bn68}, multiple spiral arms \citep{mwc53} and an outer warped disk \citep{kerr57}.  
Because we are located deep within the Galactic plane,
many attempts to understand its precise morphology and kinematics remain restricted \citep[for reviews, see e.g.,][]{fc10,xhw18,sz20}.

In recent decades, the optical data releases of the {\it Gaia} mission \citep{gaia18,dr323} and developments of high-precision radio Very Long Baseline Interferometry (VLBI) techniques \cite[e.g.,][]{xrzm06,hbm+06,brm+11,rh14} have provided unique opportunities to map the Galactic structure  and analyze its kinematic signature with accurate trigonometric parallax and proper motion measurements. 
Taking advantage of the most recent {\it Gaia} Data Release 3 (DR3; \citealt{dr323}), many efforts have been made not only to update the spiral arm segments in the solar neighbourhood using  both young \citep[e.g.,][]{hxh21,xhb+21,xu+23,hou21,lxh+23,s25} and old \citep[e.g.,][]{lxh+22,po21} tracers, but also to reveal the warping signatures in the outer Galactic disk \citep[e.g.,][]{ywc+21,cnl22,he23}. Nonetheless, most of the {\it Gaia} DR3 stars with accurate parallax distances are confined to the solar neighborhood, within distances of $\sim$~5\,kpc. 
Complementary to \textit{Gaia}, VLBI surveys dedicated to mapping the Galactic disk have yielded astrometric results for over 200 high-mass star-forming region (HMSFR) masers in the northern hemisphere, the majority of which are concentrated within $\sim$~10\,kpc on the near side and account for over a third of the Galactic disk \citep{reid19,vera20,xu+23}. 

However, mapping the far side of the Milky Way beyond the Galactic center remains a critical challenge \citep{reid24}. 
This region, also referred to as part of the Extreme Outer Galaxy (i.e., outside the Outer Arm or at Galactocentric radii $R \geq 2R_0$, where $R_0$ denotes the Galactocentric distance of the Sun; \citealt{ddt94}), has long been difficult to study owing to gas deficiency, far distances, and line-of-sight obscuration.
Despite this, \cite{dt11} discovered a new molecular spiral arm beyond the Outer Arm in the first Galactic quadrant at $R\sim$ 15\,kpc, later named Outer Scutum$-$Centaurus (OSC) Arm as an extension of Scutum$-$Centaurus Arm. Over the past decade, the Milky Way Imaging Scroll Painting (MWISP; \citealt{syz+19,syy+21,mwisp25}) project has carried out the deepest CO survey to date, achieving high sensitivity and extensive velocity/Galactic latitude coverage. This survey has revealed that the OSC Arm extends over 40\,kpc from the first to the third Galactic quadrant, possibly making it the longest spiral arm in the Milky Way \citep{sxy+15,ssz+17,syz+24}.
Nevertheless, distances to these molecular clouds were derived using traditional kinematic methods.

So far, there is a lack of accurate distance measurements to the far side of the Galaxy.  \cite{srd+17} first reported a star-forming region G007.47+0.06 in this area, with a trigonometric parallax distance of 20\,$\pm$\,2\,kpc, though this almost approaches the best achievable uncertainty for current VLBI arrays. At the same time, an updated method called three-dimensional (3D) kinematic distance was developed, which incorporates both Local Standard of Rest (LSR) velocities and proper motions to yield more accurate distance estimates \citep{yyh+16}.  Based on Bayesian inference, this method combines the likelihoods of these velocity components relative to a Galactic rotation model. By exploiting the geometric constraints between velocity and distance, it infers a robust distance at the peak of the combined 3D posterior distribution. Subsequently, \cite{reid22} further evaluated the accuracy and feasibility of 3D kinematic distance, demonstrating its superiority over direct parallax measurements for sources at distances $\gtrsim$ 8 kpc. More recently,  \cite{szx+23} reported an HMSFR G034.84-00.95 at a 3D kinematic distance of $\sim$~19\,kpc.

In this work, we present VLBI astrometric observations to derive the proper motions and 3D kinematic distance for a 22\,GHz water maser associated with the star-forming region G040.96+02.48. This source lies in the OSC Arm on the far side of the Galaxy and is associated with \textit{WISE} \hii\ region G040.954+02.473 \citep{aaj+15} and the molecular cloud MWISP G040.958+02.483 \citep{ssz+17}. \cite{aab+17} carried out radio continuum observations toward the $\hii$ region G040.954+02.473 and searched for other molecular line species, but only H$_2$O maser emission was detected. Meanwhile, \cite{sxc+18} detected a new CH$_3$OH maser and identified a young stellar object in the molecular cloud G040.958+02.483. %, in addition to the known H$_2$O maser.
These observations collectively indicate ongoing massive star formation.

The remainder of this paper is organized as follows:
Section~\ref{sec:observations} describes the VLBI observations and data reduction. The astrometric results are presented in Section~\ref{sec:results}, followed by discussion in Section~\ref{sec:discussion}. A summary is given in Section~\ref{sec:summary}.

\section{Observations} \label{sec:observations}
We carried out 10 epochs of VLBI observations of the 22\,GHz H$_2$O maser toward the star-forming region G040.96+02.48 between 2021 November 14 and 2023 January 13, using the East Asian VLBI Network (EAVN; \citealt{asi18,eavn22}). The EAVN currently comprises arrays from three constituent networks: the VLBI networks in Japan (actually including JVN and VERA), the Chinese VLBI Network (CVN), and the Korean VLBI Network (KVN)\footnote{\url{https://radio.kasi.re.kr/eavn/main.php}}. It operates primarily at 6.7, 22, and 43\,GHz, achieving corresponding angular resolutions of 1.82, 0.55, and 0.28 mas with a maximum baseline length of 5,078\,km. We observed the maser source at ($\alpha,\delta)_{\rm J2000.0}$ = (18:57:14.5900, 08:15:54.800) and the phase reference J1856+0610 at ($\alpha,\delta)_{\rm J2000.0}$ = (18:56:31.8388, 06:10:16.765) for relative VLBI astrometry. The separation angle between the target maser and the phase reference is 2.1\degr.  The observational details for this work, including the observing epochs and participating antennas, are listed in Table\,\ref{tab:tableb1}. Also, the epochs without “geodetic” blocks \footnote{To compensate for tropospheric systematic errors, we incorporated “geodetic” blocks into each observing session. This technique follows standard geodetic VLBI procedures by performing rapid, successive observations of $\sim$ 10 calibrators across a wide range of azimuths and elevations to obtain broadband group delay measurements. This approach enables us to solve for instrumental clock offsets and residual atmospheric path-delays with centimeter-level precision, thereby significantly enhancing the overall astrometric accuracy \citep{rmb+09,reid22}.} are specifically indicated. The observation modes of each network, as well as the raw data recording and correlation procedures, have been described in detail by \cite{szx+23}. Thus, they are not repeated here.

The correlated raw data were reduced with the NRAO Astronomical Image Processing System (AIPS; \citealt{aips+96,greisen03}) and ParselTongue software \citep{kvr+06}. For epochs that included geodetic blocks, we performed separate reductions with and without applying these geodetic blocks. The data reduction generally followed the standard procedures as described in \cite{rmb+09}. We note that for the Tianma telescope, a large delay occurred at epoch A2221c and no solution was found at epoch A2221d during the manual phase calibration in the process of phase referencing. Accordingly, we excluded the Tianma telescope from imaging for these two epochs (see Table\,\ref{tab:tableb1}). Additionally, high system temperatures occurred at epoch A2221e. The final epoch was excluded from further processing due to its excessively high system temperature.
After data reduction for each epoch, the positional offsets of maser spots in the eastward and northward directions
were recorded for subsequent analysis. 
Only a single discernible and sufficiently compact maser spot was detected across all epochs, peaking at LSR velocity $\sim$~$-$54\,km~s$^{-1}$. 
To demonstrate the robustness of our results, the positional offsets derived from both reduction pipelines (with and without geodetic blocks) are tabulated in Table\,\ref{tab:tableb1} for comparison.

\rotate
\begin{deluxetable*}{llllccccccccccc}
\tablecaption{Observational information of star-forming region G040.96+02.48}
\tablewidth{\textwidth}
\tablecolumns{14}

\tablehead{
\colhead{} & \colhead{} & \colhead{} & \colhead{} &
\multicolumn{5}{c}{Without geodetic blocks} & \colhead{} &
\multicolumn{5}{c}{With geodetic blocks} \\
\cline{5-9} %
\cline{11-15}
\colhead{Epoch} & \colhead{MJD} & \colhead{Antenna} & \colhead{$\lambda$} &
\colhead{$\Delta x$} & \colhead{$\Delta y$} & \colhead{F$_{\rm peak}$} & \colhead{S/N} & \colhead{Beam} &\colhead{}&
\colhead{$\Delta x$} & \colhead{$\Delta y$} & \colhead{F$_{\rm peak}$} & \colhead{S/N} & \colhead{Beam} \\
\colhead{} & \colhead{} & \colhead{} & \colhead{(km)} &
\colhead{(mas)} & \colhead{(mas)} & \colhead{(Jy beam$^{-1}$)} & \colhead{} & \colhead{(mas$\times$mas@$^\circ$)} & \colhead{} &
\colhead{(mas)} & \colhead{(mas)} & \colhead{(Jy beam$^{-1}$)} & \colhead{} & \colhead{(mas$\times$mas@$^\circ$)}
}

\setlength{\tabcolsep}{2pt}

\startdata
{\bf A2138a} & 59532 & 1234(5) & 700 & 0.164$\pm$0.097 & 13.290$\pm$0.053 & 0.85$\pm$0.03 & 24 & 5.4$\times$2.8@78& & 2.386$\pm$0.128 & -0.264$\pm$0.072 & 0.72$\pm$0.04 & 18 & 5.4$\times$2.7@75 \\
{\bf A2201a} & 59615.5 & (2)34567 & 4500 & -1.339$\pm$0.080 & 7.700$\pm$0.077 & 1.40$\pm$0.13 & 11 & 2.5$\times$0.9@134& & \nodata & \nodata & \nodata & \nodata & \nodata \\
A2201b & 59645.5 & 1246(7)8 & 1800 & 0.891$\pm$0.046 & -1.344$\pm$0.061 & 1.07$\pm$0.08 & 14 & 2.2$\times$1.1@33& & 0.516$\pm$0.017 & -1.694$\pm$0.023 & 2.48$\pm$0.08 & 32 & 1.7$\times$1.0@27 \\
A2201c & 59687.5 & 1236(7) & 1300 & 0.339$\pm$0.010 & -1.131$\pm$0.014 & 5.11$\pm$0.06 & 87 & 2.7$\times$1.3@153& & 0.251$\pm$0.010 & -1.348$\pm$0.015 & 5.30$\pm$0.06 & 85 & 2.7$\times$1.3@154 \\
A2201d & 59708 & (1)234(5)6(7)8 & 1800 & 0.422$\pm$0.021 & -1.739$\pm$0.033 & 3.60$\pm$0.15 & 23 & 1.8$\times$0.9@21& & 0.386$\pm$0.022 & -1.620$\pm$0.032 & 3.76$\pm$0.16 & 23 & 1.8$\times$0.9@25 \\
{\bf A2201e} & 59742 & 234 & 500 & 1.324$\pm$0.277 & -1.559$\pm$0.130 & 3.13$\pm$0.11 & 29 & 6.5$\times$2.9@95& & \nodata & \nodata & \nodata & \nodata & \nodata \\
A2221b & 59863 & 1234(5)678(9) & 2300 & -0.545$\pm$0.022 & -2.942$\pm$0.024 & 22.9$\pm$0.12 & 187 & 1.3$\times$1.2@35& & -0.735$\pm$0.019 & -3.107$\pm$0.021 & 22.7$\pm$0.11 & 209 & 1.4$\times$1.2@33 \\
A2221c & 59888 & 123(5)6(7)(8)(9) & 1300 & -0.777$\pm$0.018 & -2.893$\pm$0.029 & 11.0$\pm$0.27 & 40 & 2.8$\times$1.3@156& & -0.730$\pm$0.015 & -3.111$\pm$0.025 & 13.4$\pm$0.28 & 48 & 2.7$\times$1.3@158 \\
A2221d & 59919 & (1)2346(7)8(9) & 1800 & -0.774$\pm$0.024 & -3.812$\pm$0.043 & 9.47$\pm$0.36 & 26 & 2.4$\times$1.2@173& & -1.174$\pm$0.024 & -3.734$\pm$0.041 & 9.94$\pm$0.38 & 26 & 2.3$\times$1.3@175 \\
{\bf A2221e} & 59957 & 1234589 & \nodata & \nodata & \nodata &  \nodata & \nodata & \nodata &  & \nodata & \nodata & \nodata & \nodata & \nodata \\
\enddata
\begin{flushleft}\noindent\textbf{Notes.} Columns\,1 and 2 give the observational epoch and Modified Julian Day (MJD), respectively. The epochs with outlier values and the final epoch are highlighted for clarity. Column\,3 lists the participating antennas (1 = Iriki; 2 = Yonsei; 3 = Ulsan; 4 = Tamna; 5 = Urumqi; 6= Mizusawa; 7 = Ishigakijima; 8 = Ogasawara; 9 = Tianma). The numbers in parentheses represent the unused antennas in data reduction due to a lack of solutions during the process of phase referencing.  Column\,4 presents the longest baseline length used in imaging. Columns\,5 through 9 provide the following parameters for masers without geodetic blocks: eastward and northward positional offsets relative to the phase center at ($\alpha,\delta)_{\rm J2000.0}$ = (18:56:31.8388, 06:10:16.765), peak flux density, signal-to-noise ratio (S/N), and (naturally weighted) beam size with position angle.  The position offsets in the eastward and northward directions, $(\Delta x, \Delta y)$, are represented by $\Delta \alpha \cos \delta$ and $\Delta \delta$, respectively, where $\alpha$ and $\delta$ are the Right Ascension (R.A.) and Declination (Dec.). Columns\,10-14 summarize the same set of parameters with geodetic blocks included. 
To account for systematic errors from unmodeled atmospheric
delays, an “error floor” was added in quadrature to the formal uncertainties for all epochs. For sessions lacking geodetic blocks, the “error floor” was set to 0.134\,mas in R.A. and 0.327\,mas in Dec.; for sessions including geodetic blocks, the values were 0.151\,mas and 0.304\,mas, respectively.
\end{flushleft}

\label{tab:tableb1}
\end{deluxetable*}

\section{Results}\label{sec:results}

\subsection{Parallax and Proper-motion Fitting}

\begin{figure*}[htbp]    
    \centering
    \includegraphics[scale=0.58]{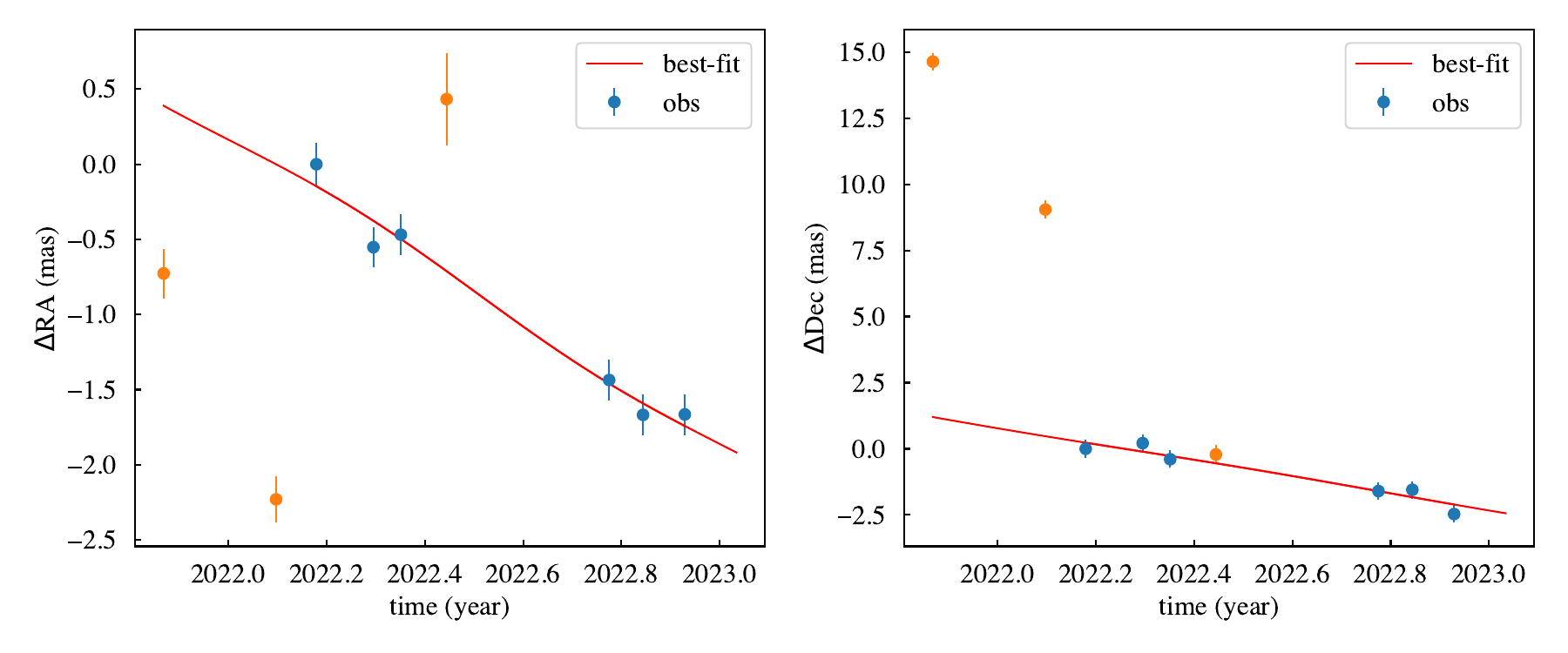}
    \includegraphics[scale=0.58]{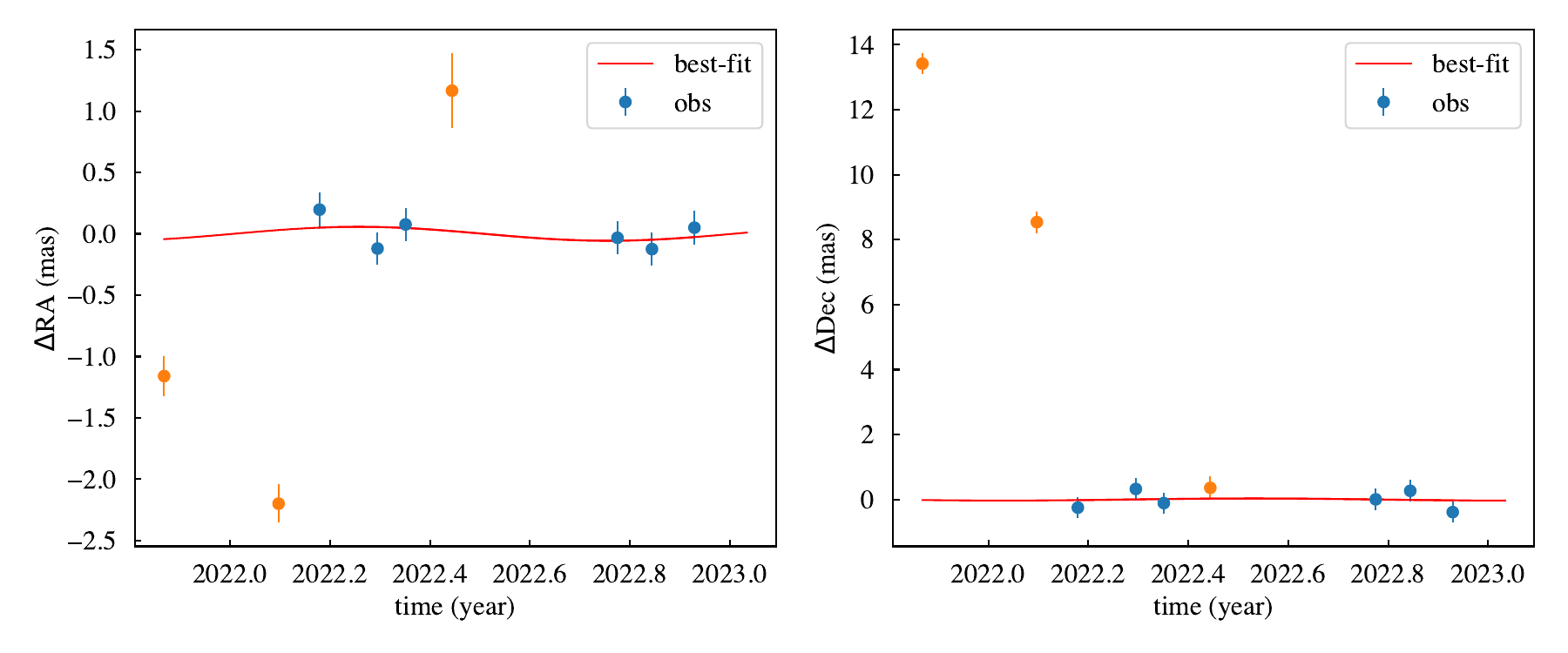}
    \caption{Parallax fitting results without incorporating  geodetic blocks. The top panels present the eastward (left) and northward (right) positional offsets as a function of time. Similar to the top panels, the bottom panels show the eastward and northward offsets with the fitted proper motion removed. Outliers are marked by orange scatters.}
    \label{fig:plxpm1}
\end{figure*}
\begin{figure*}[htbp]    
    \centering
    \includegraphics[scale=0.58]{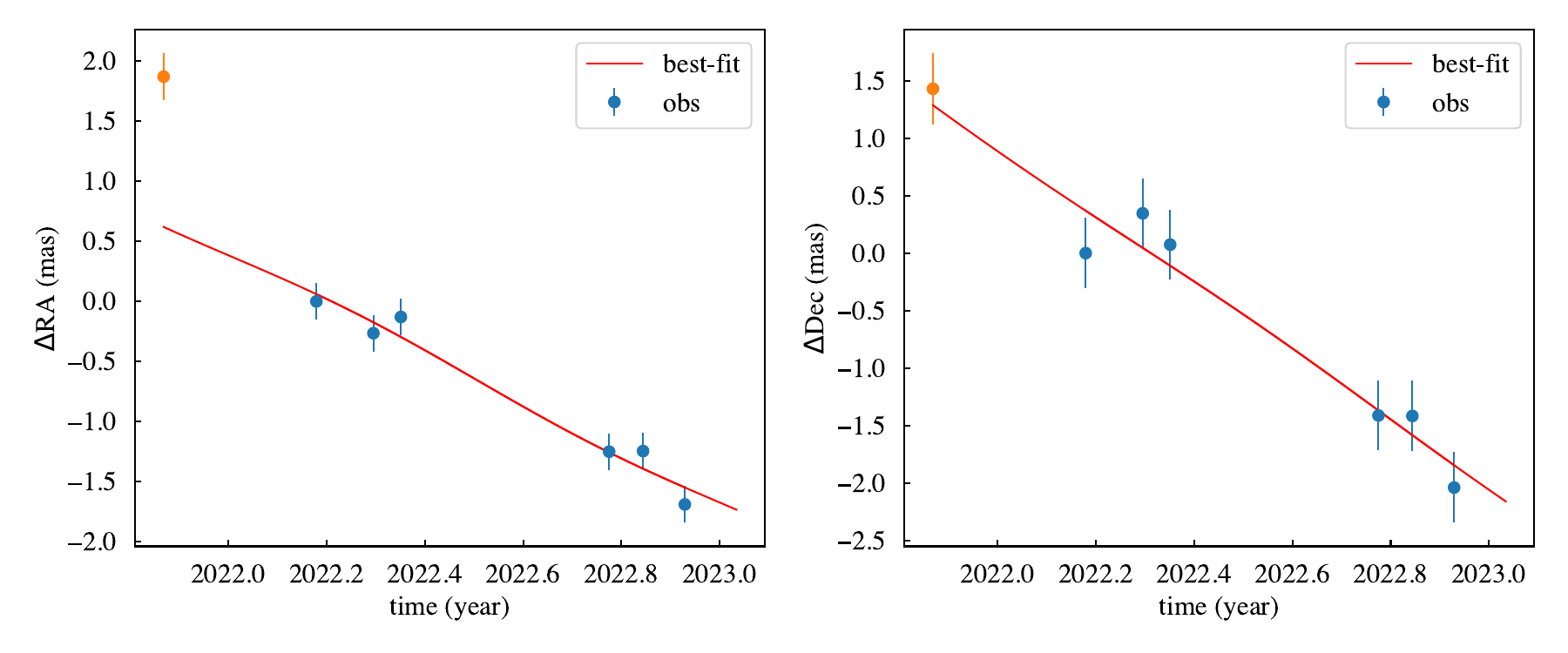}
    \includegraphics[scale=0.58]{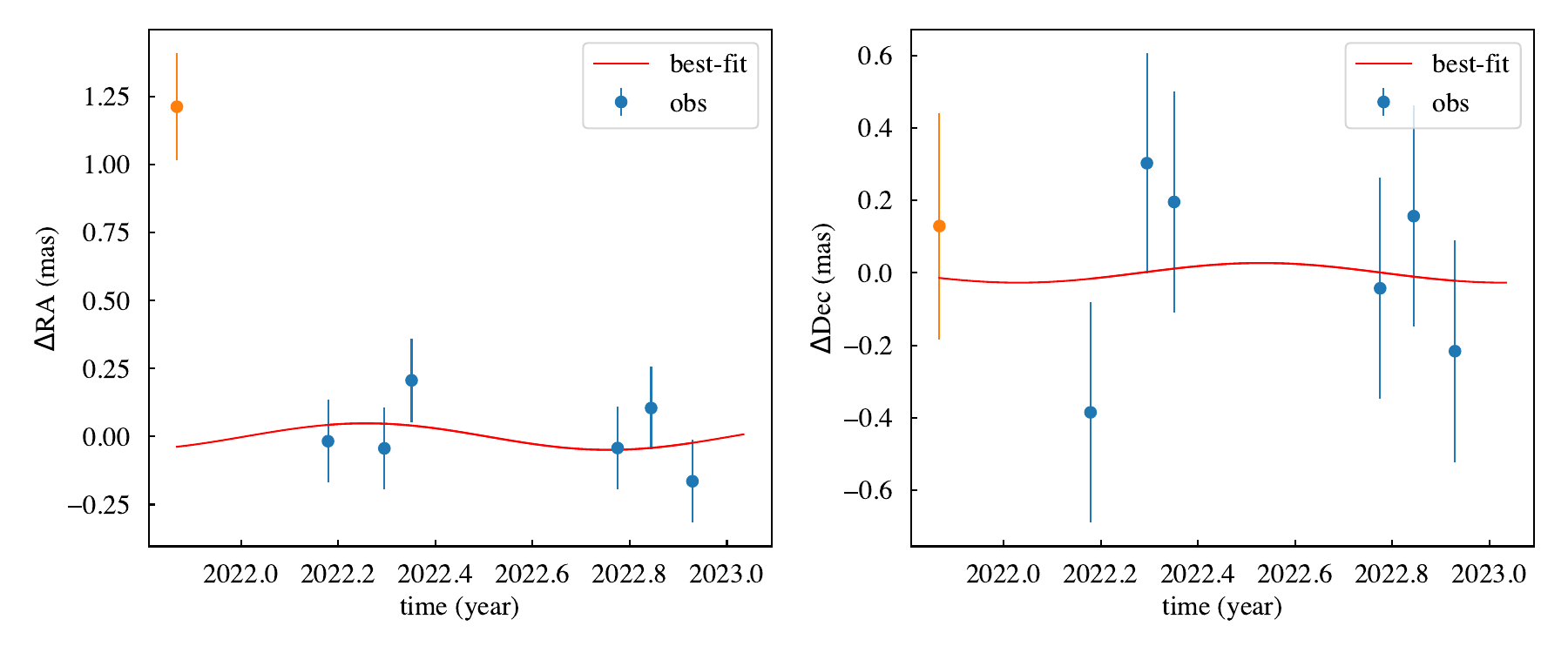}
    \caption{Same as Fig.\,\ref{fig:plxpm1} but with consideration of geodetic blocks.}
    \label{fig:plxpm2}
\end{figure*}
Prior to fitting, we introduced “error floors” to account for systematic position uncertainties. These were added in quadrature to the formal uncertainties in the eastward and northward directions, yielding a reduced $\chi^2$ per degree of freedom close to unity \citep[e.g.,][]{xbr+21,bwx+24}. After this step, we used the Markov Chain Monte Carlo method to obtain the best-fit results, performing separate fits with and without the geodetic blocks. As shown in both  Figs.\,\ref{fig:plxpm1} and \ref{fig:plxpm2}, an obvious outlier appears at the first epoch.  Moreover, the second and sixth epochs (i.e., A2201a and A2201e) without geodetic blocks  show larger deviations in Fig.\,\ref{fig:plxpm1}. These anomalies are likely attributable to the use of dual-beam mode at the first two epochs. The sixth epoch included only three KVN antennas and exhibited phase jumps, and we consequently removed it from the analysis. After excluding these outlier epochs, the fits yield consistent results. The best-fit parameters are summarized in  Table\,\ref{tab:table1}. 
We do not obtain a sufficiently reliable parallax measurement for the star-forming region G040.96+02.48, similar to the case of G034.84$-$00.95 \citep{szx+23}. Finally, the case with geodetic blocks is retained for subsequent analysis.

\begin{table}
    \centering
    \caption{Parallax and proper motion measurements for star-forming region G040.96+02.48 }
    \begin{tabular}{lccc}
      \hline
      \hline
       Group &  $\pi$ & $\mu_{\alpha}$cos$\delta$ & $\mu_{\delta}$\\
       & (mas) & (mas~yr$^{-1}$) & (mas~yr$^{-1}$)\\
      \hline
     Without geodetic  & 0.06$\pm$0.17 & $-2.01_{-0.48}^{+0.50}$ & $-3.12_{-0.47}^{+0.46}$\\
     With geodetic & $0.05\pm0.18$ & $-2.06_{-0.51}^{+0.53}$ & $-2.95_{-0.44}^{+0.45}$\\
      \hline
    \end{tabular}
    \label{tab:table1}
\end{table}

\subsection{3D Kinematic Distance and Peculiar Motion Estimates}
\begin{figure}
    \centering
    \includegraphics[width=1.0\linewidth]{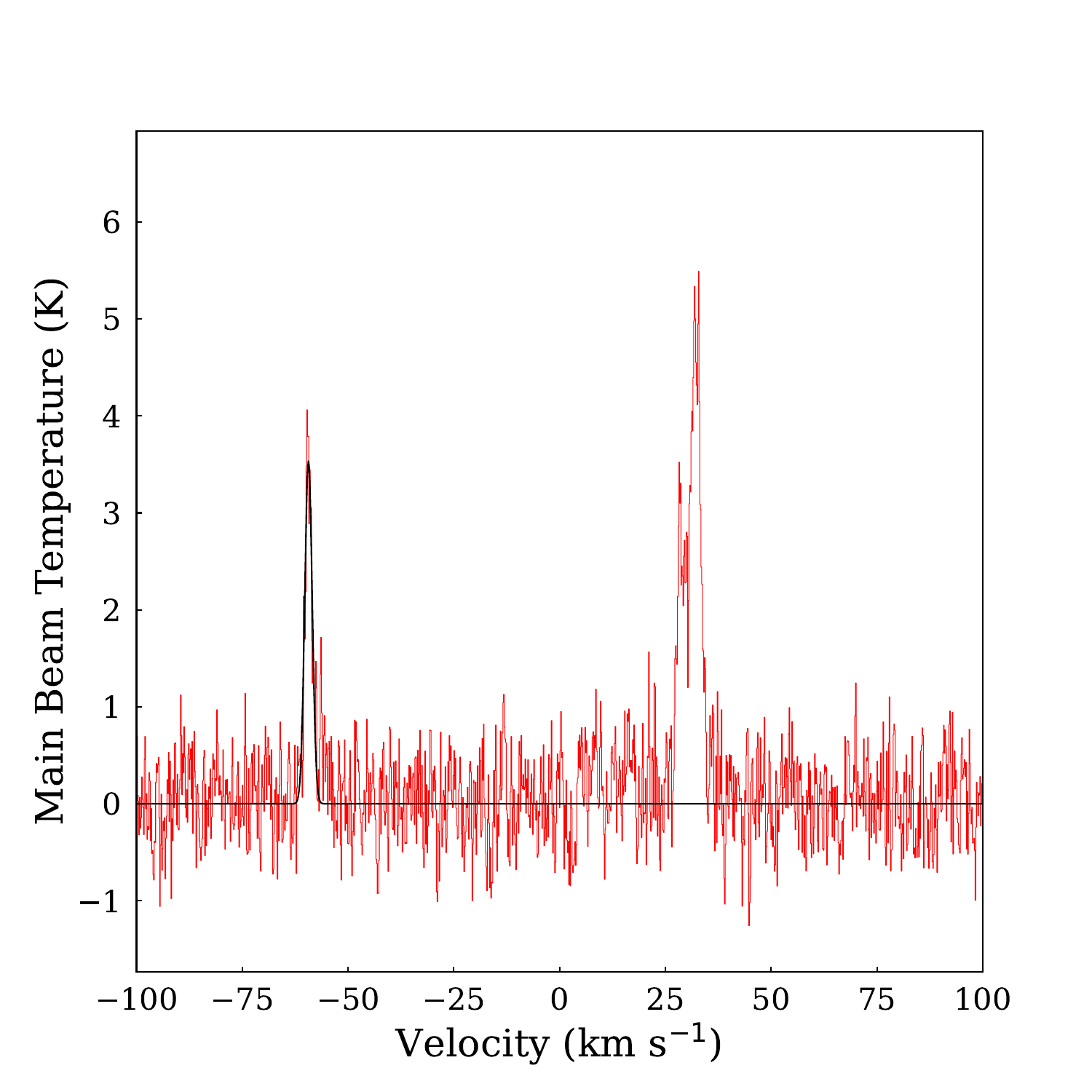}
    \caption{$^{12}$CO($J$=1-0) spectrum of  the star-forming region G040.96+02.48. The spectral profile is presented by the red curve, with a Gaussian fit to the far component overlaid as a black line. The LSR velocity component associated with the far side peaks at $-$59.4\,km~s$^{-1}$.}
    \label{fig:co}
\end{figure}

\begin{table*}
    \centering
    \caption{Summarized physical parameters for star-forming region G040.96+02.48 }
    \begin{tabular}{lccccccc}
      \hline
      \hline
       Source & $d_{\rm 3D}$& $d_{\rm 1D}$ & {\it z} & $V_{\rm LSR}$ & $U_{\rm s}$ & $V_{\rm s}$ & $W_{\rm s}$\\
       & (kpc) & (kpc) &(pc)& (km~s$^{-1}$) & (km~s$^{-1}$)& (km~s$^{-1}$)& (km~s$^{-1}$)\\
      \hline
      G040.96+02.48 & 20.2$\pm$3.2 & 17.7$\pm$0.7 & 872$\pm$139  & $-$59$\pm$5 &$-$32$\pm$18 & $-$42$\pm$73&52$\pm$51\\ 
     
      \hline
    \end{tabular}
    
    \noindent\textbf{Notes.} \raggedright{ $d_{\rm 3D}$ indicates the 3D kinematic distance. The updated traditional kinematic distance ($d_{\rm 1D}$) is also provided for comparison.  $z$ denotes the vertical height derived from the 3D kinematic distance. The LSR velocity ($V_{\rm LSR}$) information comes from the associated $^{12}$CO molecular cloud. The source's peculiar motion is decomposed into three orthogonal components: $U_{\rm s}$, $V_{\rm s}$ and $W_{\rm s}$, which are directed radially toward the Galactic Center, tangentially in the direction of Galactic rotation, and vertically toward the North Galactic Pole, respectively (see Figure 8 in \cite{rmz+09} for detailed geometry).} 
    \label{tab:table2}
\end{table*}
\begin{figure}
    \centering
    \includegraphics[width=1.0\linewidth]{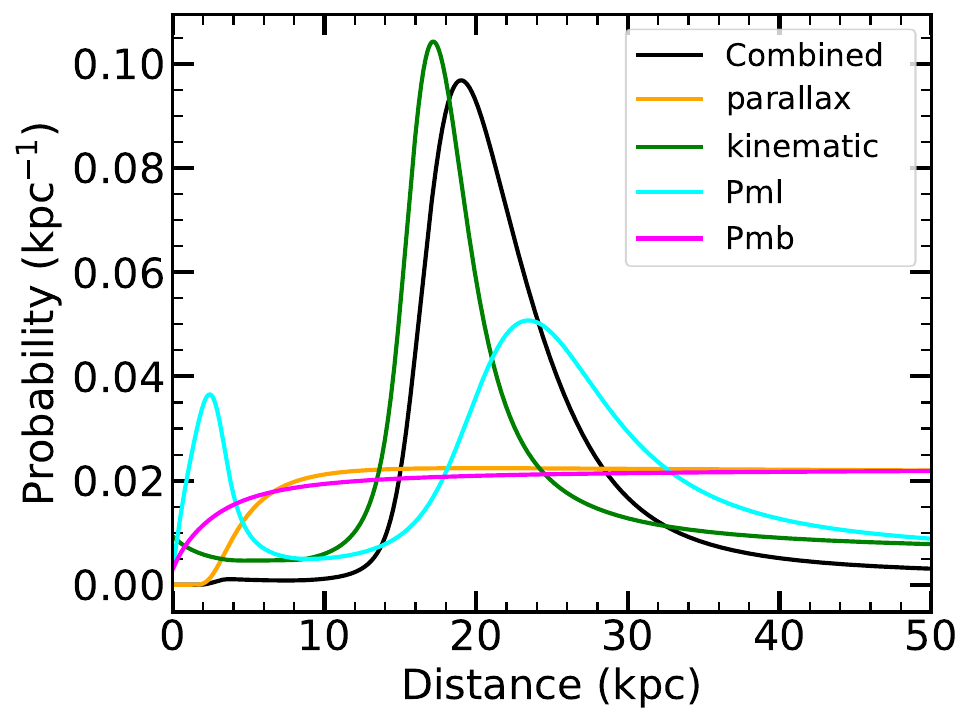}
    \caption{Distance PDFs for star-forming region G040.96+02.48. The term "kinematic" denotes the PDF derived solely from the LSR velocity. Abbreviations for the proper motions in Galactic longitude and latitude directions are Pml and Pmb, respectively. The black curve combines different components (colored curves) and yields a 3D kinematic distance of 20.2$\pm$3.2\,kpc. }
    \label{fig:3dkd}
\end{figure}
\subsubsection{LSR Velocity of G040.96+02.48}
The 3D kinematic distance is determined using the source's full kinematic information, including both the proper motion discussed above and the LSR velocity, as the latter can be more complex to ascertain. \cite{ssz+17} have reported the associated molecular cloud G040.958+02.48 with a LSR velocity of $-$59.35\,km~s$^{-1}$, but the water maser exhibits multi-component velocities spanning from $-$62 to $-$42\,km~s$^{-1}$ in previous observations \citep[e.g.,][]{aab+17,sxc+18}. Notably, the new CH$_3$OH maser detected by \cite{sxc+18} exhibits a single component peaking at a LSR velocity $\sim$ $-$62.5\,km~s$^{-1}$. We inspected the MWISP $^{12}$CO spectrum to further confirm the velocity component and found that there was only one prominent peak with negative velocity comparable to maser observations (see Fig.\,\ref{fig:co}). We therefore assume the molecular cloud velocity is associated with the masers, adopting an uncertainty of 5\,km~s$^{-1}$ \citep[e.g.,][]{hbm+06,hna+12}. 
\subsubsection{Derivation of 3D Kinematic Distance}
We then applied the 3D kinematic distance technique, which combines multiple probability density functions (PDFs) derived from the full 3D kinematic information and incorporates the updated Galactic rotation curve and solar motion parameters (i.e., fit A5; \citealt{reid19}), to the star-forming region G040.96+02.48. Fig.\,\ref{fig:3dkd} displays the resulting PDFs for our target source, and the 3D kinematic distance is finally placed at 20.2$\pm$3.2\,kpc. This value is larger than the traditional kinematic distance estimate ($17.7$ kpc), likely due to the constraints imposed by the proper motion data. 
\subsubsection{3D Galactic Location of G040.96+02.48}
We derived a vertical height of 872\,pc from the 3D kinematic distance measurement, indicating a pronounced warp in the outer Galactic disk. This value lies within the vertical height range
(i.e., $-500$--$1500$\,pc) documented in \cite{ssz+17} for molecular clouds in the Extreme Outer Galaxy, spanning Galactic longitudes from $34.75^{\circ}$ to $45.25^{\circ}$. In Fig.\,\ref{fig:mw}, the star-forming region G040.96+02.48 is shown projected onto the Galactic disk, where it appears slightly outside the OSC Arm at {\it R}~=~15\,kpc. Also included for comparison are two other star-forming regions located on the far side.
The spatial distribution of the three HMSFR masers is broadly consistent with the commonly adopted spiral model. However, the limited number of sources with precise distances currently prevents us from placing strong constraints on the morphology and location of the OSC Arm in the first Galactic quadrant.

\subsubsection{Peculiar Motions of G040.96+02.48}
\begin{figure*}
    \centering
    \includegraphics[width=0.85\linewidth]{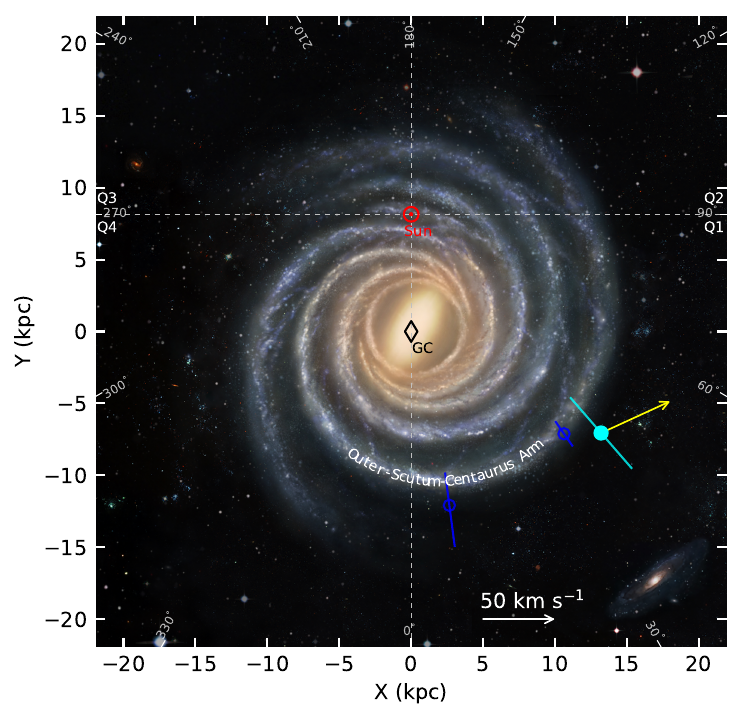}
    \caption{Spatial distributions of three most distant star-forming regions with precise distance measurements projected onto the Galactic plane. The sources shown in clockwise order are G040.96+02.48 (this work), G034.84$-$00.95 \citep{szx+23}, and G007.47+0.06 \citep{srd+17}. The yellow arrow shows the peculiar motion vector of star-forming region G040.96+02.48. A velocity scale of 50 km~s$^{-1}$ is indicated in the plot. The background is a new artist's impression of the Milky Way credited by Xing-Wu Zheng $\&$ Mark Reid BeSSeL/NJU/CFA. The Sun's position is at (0, 8.15)\,kpc, while the Galactic Center is at (0.0, 0.0)\,kpc.}
    \label{fig:mw}
\end{figure*}

Using the full kinematic information and distance estimate, we derived the target's peculiar (non-circular) motions following \cite{rmz+09} and \cite{xbr+21}, adopting the revised Galactic parameters and solar motions (i.e., fit A5) from \cite{reid19}. The resulting physical parameters are summarized in Table\,\ref{tab:table2}. We note that the peculiar motions in all three directions exhibit large values with considerable uncertainties. The other two star-forming regions on the far side also show similar behavior.  In Fig.\,\ref{fig:mw}, the resulting peculiar motion is shown projected onto the Galactic plane (see the yellow vector). The in-plane velocity vector exceeds 50\,km~s$^{-1}$ and shows a substantial deviation from the direction of Galactic rotation. The most prominent feature is a relatively well-constrained outward radial motion ($U_{\rm s}$) of $-32\pm$18\,km~s$^{-1}$. In particular, the component in the direction of Galactic rotation ($V_{\rm s}$) exhibits a large lag, substantially greater than the typical $\sim$~15 km s$^{-1}$ lag relative to the Galactic rotation curve \citep{rmz+09}.  Furthermore, the observed significantly large vertical motion ($W_{\rm s}$) may be related to the Galactic warp. These kinematic signatures suggest that the dynamical behavior in the outer Galaxy may be comparable in complexity to that observed near the end of the Galactic bar \citep{reid19}.
\section{discussion}\label{sec:discussion}
The Galactic disk is known to exhibit significant warping, which begins at a Galactocentric radius of roughly 7 to 8\,kpc. This large-scale distortion  manifests as an S-shaped structure, with its outer disk bent upward in the north and downward in the south, as widely revealed by a variety of gaseous \citep[e.g.,][]{burke57,kerr57,gkw60,lbh06,sxy+15,ssz+17}, stellar \citep[e.g.,][]{lcg+02,mzg+06,rma+19,pdl+18,cwd+19,ssm+19,pda+20} and dusty  \citep[e.g.,][]{ds01,dcl03,mrr+06} tracers. 

The most recent characterization of the Galactic warp is provided by a precessing warp model based on the 3D distributions of classical Cepheids \citep[e.g.,][]{cwd+19,ssm+19}. Classical Cepheids are young stars and are thus expected to preserve the warp signature of their birthplaces in the Galactic disk. In this context, very young star-forming regions can be used to assess the precessing warp model. G040.96+02.48 appears more representative of the warped outer Galactic disk than the other two sources: G007.47+00.05 lies near the warp's line of nodes at $R\sim12$\,kpc with a vertical height of only 18\,pc (see Fig.\,3 of \citealt{cwd+19}), whereas G034.84$-$00.95 is located at negative Galactic latitude \citep{ssz+17,szx+23}. To quantitatively evaluate the precessing warp model, we applied their power-law model $z_w = a (R - R_w)^b \sin(\phi - \phi_w)$ to G040.96+02.48, adopting the best-fit parameters from the combined optical and {\it WISE} samples (see Table\,1 in \cite{cwd+19} for details). This model predicts a maximum warping amplitude of 841\,pc  (i.e., assuming $\sin(\phi - \phi_w)=1$) and a value of 717\,pc at $\ell=40.96^{\circ}$, both of which are in good agreement with the observed vertical height of 872\,pc. This result is also consistent with the findings of \cite{pkd+25}, who reported the latest warp amplitude of $\sim$ 700\,pc at $R\sim$ 14\,kpc using {\it Gaia} DR3 young giants and classical Cepheids.
\begin{figure*}
 \centering
    \includegraphics[width=0.99\linewidth]{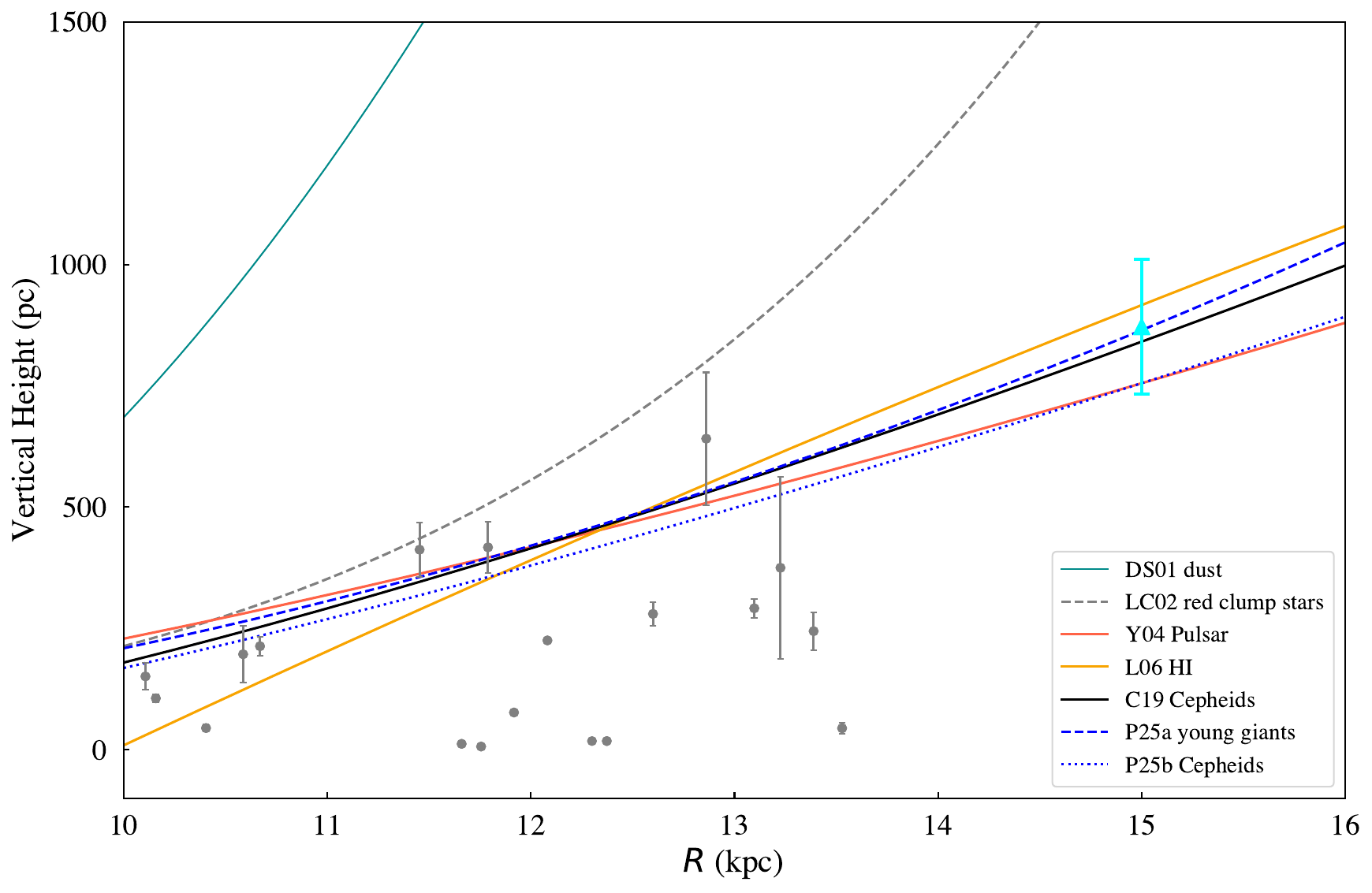}
    \caption{Maximum amplitudes of the northern warp overlaid with the vertical heights of HMSFR masers. Similar to \cite{cwd+19}, all maser samples are presented merely to illustrate a simple warping trend. The gray dots indicate the HMSFR masers collected from the literature, while the cyan triangle marks G040.96+02.48. The curves show maximum warp amplitudes derived from different studies: DS01, dust \citep{ds01}; LC02, red clump stars \citep{lcg+02}; Y04, pulsars \citep{y04}; L06, $\hi$ \citep{lbh06}; C19, Cepheids \citep{cwd+19}; P25a (young giants) and P25b (Cepheids) \citep{pkd+25}.}
    \label{fig:wp}    
\end{figure*}

To further investigate the Galactic warp in the northern hemisphere using star-forming regions with well-determined distances, we collected additional star-forming regions from the literature \citep[e.g.][and references therein]{reid19,vera20,xu+23,bwx+24}, with the criteria being reliable trigonometric parallax measurements and $R >$ 10\,kpc. Their vertical heights, together with our target source, were overlaid on the predicted maximum warp amplitude as a function of $R$ derived from several classical or recent warp models.
As shown in Fig.\,\ref{fig:wp}, these star-forming regions with accurate distances trace the blurry trend of the Galactic warp, given the limited sample size currently available. As the most distant star-forming region with a reliably measured distance to date, our source is consistent with the moderately rising warp models (e.g., Cepheids and $\hi$ tracers).

\section{Summary}\label{sec:summary}
In this paper, we present VLBI astrometric results of a 22\,GHz water maser associated with the star-forming region G040.96+02.48 on the far side of the Milky Way, conducted with the EAVN. The main results are summarized as follows:

(1) Using astrometric data from six reliable epochs, we derived a proper motion of {\bf ($\mu_{\alpha}\cos\delta, \mu_{\delta}$) = ($-2.06_{-0.51}^{+0.53}$,  $-2.95_{-0.44}^{+0.45}$)}~mas~yr$^{-1}$ for the target water maser.

(2) Based on full 3D kinematic information, the star-forming region G040.96+02.48 is placed at a distance of 20.2$\pm$3.2\,kpc, slightly outside the OSC Arm.

(3) The star-forming region G040.96+02.48 exhibits a large vertical height of $\sim$ 872\,pc based on the 3D kinematic distance estimate, which is in good agreement with the prediction of the precessing warp model.

(4) The peculiar motions of G040.96+02.48 are complex. In particular, the source shows an outward motion of $-32\pm$18\,km~s$^{-1}$ relative to the radial direction. 
Note that the remaining two velocity components are highly uncertain and should therefore be treated with caution.

Overall,  we have reported the astrometric results for a 22\,GHz H$_2$O maser associated with star-forming region G040.96+02.48 on the far side. This source exhibits an obvious Galactic warp feature and complex 3D kinematics. Our observations substantially expand the sample of star-forming regions with accurate distances in the Extreme Outer Galaxy. Subsequent work will be dedicated to extending survey coverage in the Extreme Outer Galaxy, aiming to constrain the morphology of the OSC Arm better and to investigate the warp and kinematic properties on the far side.

\begin{acknowledgments}
We sincerely thank the anonymous referee for suggestions which greatly improved the manuscript. 
This work was funded by the National Key RD Program of China (grant No.2024YFA1611504), the NSFC Grants 12403077, 12303072, 12403041, 12503071, the National SKA Program of China (grant No.2022SKA0120103), the Xinjiang Talent Development Fund (grant No.XJRC-2025-KJ-YJ-CXPT-180) and the Key Laboratory for Radio Astronomy. This work is made use of the East Asian VLBI Network (EAVN), which is operated under cooperative agreement by National Astronomical Observatory of Japan (NAOJ), Korea Astronomy and Space Science Institute (KASI), Shanghai Astronomical Observatory (SHAO), Xinjiang Astronomical Observatory (XAO), Yunnan Observatories (YNAO), National Geographic Information Institute (NGII), and National Astronomical Research Institute of Thailand (Public Organization) (NARIT), with the operational support by Ibaraki University, Yamaguchi University, and Kagoshima University. This research made use of the data from the Milky Way Imaging Scroll Painting (MWISP) project, which is a multi-line survey in $^{12}$CO/$^{13}$CO/C$^{18}$O along the northern Galactic plane with PMO-13.7m telescope. We are grateful to all the members of the MWISP working group, particularly the staff members at PMO-13.7m telescope, for their long-term support. MWISP was sponsored by National Key R$\&$D Program of China with grants 2023YFA1608000 $\&$ 2017YFA0402701 and by CAS Key Research Program of Frontier Sciences with grant QYZDJ-SSW-SLH047.
\end{acknowledgments}

\facility{EAVN}
% \software{emcee \citep{emcee}}
\vspace{5mm}

\bibliography{sample631}{}
\bibliographystyle{aasjournal}

% \allauthors

\end{CJK*}
\end{document}